\documentstyle[epsfig]{mn}
\def\Msun{M_{\odot}}
\def\simlt{\mathrel{\rlap{\lower 3pt\hbox{$\sim$}}\raise 2.0pt\hbox{$<$}}}
\def\simgt{\mathrel{\rlap{\lower 3pt\hbox{$\sim$}} \raise 2.0pt\hbox{$>$}}}
\def\lta{\mathrel{\rlap{\lower 3pt\hbox{$\sim$}}\raise 2.0pt\hbox{$<$}}}
\def\gta{\mathrel{\rlap{\lower 3pt\hbox{$\sim$}} \raise 2.0pt\hbox{$>$}}}

\def\Msun{M_{\odot}}

\def\Xback{\mbox{erg s$^{-1}$ cm$^{-2}$ deg$^{-2}$}}
\def\Xflux{\mbox{erg s$^{-1}$ cm$^{-2}$}}

\begin{document}

\title[Unresolved X-ray Background]{Unresolved X-ray background: clues on galactic nuclear activity at $z>6$}
\author[Salvaterra, Haardt, Volonteri]{Ruben Salvaterra$^1$, Francesco Haardt$^{1,3}$, Marta Volonteri$^{2,3}$\\
$1$ Dipartimento di Fisica e Matematica, Universit\'a dell'Insubria,
Via Valleggio 11, 22100 Como, Italy\\
$2$ Institute of Astronomy, Madingley Road, Cambridge CB3 0HA, UK\\
$3$ Kavli Institute for Theoretical Physics, University of California
Santa Barbara, CA 93106, USA
}

\maketitle \vspace {7cm}

\begin{abstract}
We study, by means of dedicated simulations of massive black hole build-up, 
the possibility to constraint the existence and nature of the AGN population
at $z\simgt6$ with available and planned X--ray and near infrared space telescopes. We find that X--ray
deep--field observations can set important constraints to the faint--end of 
the AGN luminosity function at very high redshift. Planned X--ray telescopes
should be able to detect AGN hosting black holes with masses down to $\simgt 10^5\;\Msun$ (i.e., X--ray luminosities in excess of $10^{42}$ erg s$^{-1}$), 
and can constrain the evolution of the population of massive black hole at early times ($6\simlt z\simlt 10$). 
We find that this population of AGN should contribute substantially ($\sim 25$\%)
to the unresolved fraction of the cosmic X--ray background in the 0.5--10 keV range, and that a 
significant fraction ($\sim 3-4$\%) of the total background intensity would remain 
unaccounted even after future X--ray observations. As byproduct, we compute 
the expected UV background from AGN at $z\simgt6$ and we discuss the possible
role of AGN in the reionization of the Universe at these 
early epochs, showing that AGN alone can provide enough ionizing photons only in the (improbable) case 
of an almost completely homogeneous inter--galactic medium. 
Finally, we show that super--Eddington accretion, suggested by 
the observed QSOs at $z\simeq 6$, must be a very rare event, 
confined to black holes living in the highest density peaks. 

\end{abstract}


\section{Introduction}

The formation of black hole (BH) seeds and their evolution have been the subject of several theoretical investigations. The "flow-chart" presented Rees (1978) still stands as a guideline for the possible paths leading to formation of BH seeds in the center of galactic structures. One possibility is the direct formation of a massive BH (MBH) from a collapsing gas cloud (Haehnelt \& Rees 1993;  Loeb \& Rasio 1994; Eisenstein \& Loeb 1995; Bromm \& Loeb 2003; Koushiappas, Bullock \& Dekel 2004;  Begelman, Volonteri \& Rees 2006). The gas can condenses to form a central massive object. The mass of the seeds predicted by different models vary, but typically are in the range $M_{BH} \sim 10^4-10^6\,M_\odot$. Alternatively, the seeds of BHs can be associated with the remnants of the first generation of stars, formed out of zero metallicity gas. The first stars are believed to form at $z\sim 20$ in halos which represent high--$\sigma$ peaks of the primordial density fie
 ld. 
The absence of metals might lead to a very top--heavy initial stellar mass function, and in particular to the production of very massive stars with masses $>100 M_\odot$ (Carr, Bond, \& Arnett 1984). If very massive stars form above 260 $M_\odot$, they would rapidly collapse to BHs with little mass loss (Fryer, Woosley, \& Heger 2001), i.e., leaving behind seed BHs with masses $M_{BH} \sim 10^2-10^3\,M_\odot$ (Madau \& Rees 2001).  
The subsequent growth of BHs from these initial seeds has been investigated in the hierarchical framework,  typically associating episodes of accretion to galaxy mergers  (Haiman 2004, Yoo \& Miralda-Escud\'e 2004, Shapiro 2005; Volonteri \& Rees 2005, 2006; Lapi et al. 2006). 

Only a few observational constraints are currently available. Many are limited to the brightest sources, probing only the very bright end of the luminosity function of quasars at $z\simeq 6$ and therefore only the upper end of the massive black hole mass function (Fan et al. 2001). The observation of very luminous quasars, powered by billion solar masses BHs, at $z\approx6$ in the SDSS survey (Fan et al. 2001), implies that a substantial population of smaller accreting BHs must exist at earlier times. 
Deeper but limited constraints have been provided for more typical AGN, with $\simeq$ 2 orders of magnitude smaller MBHs, using deep X-ray observations (e.g., Alexander et al. 2001; Barger et al. 2003b; Koekemoer et al. 2004; Mainieri et al. 2005). However, these constraints lack the detail
and depth required to  understand and determine the global evolution of BH seeds and of the average MBH population at very high redshifts. Accreting BHs are observed as X--ray emitters up to $z \simlt 6$ (Barger et al. 2003, Steffen et al. 2006),  and there are no reasons to believe that higher redshift AGN would be any different. The detection of very high redshift AGN is one of the goals of 
future space missions. In particular, X--ray telescopes, such as the planned {\it XEUS}
and {\it Constellation--X}, are expected to detect these sources even at $z\simgt6$. Moreover, X--ray observatories 
can identify even heavily obscured AGN activity, due to the penetrating nature of
hard X-rays.

Additional, albeit less direct, limits on the early MBH population can already be placed by the requirement that the cumulative emission from the predicted high redshift sources does not saturate the observed unresolved X-ray background (Dijkstra, Haiman \& Loeb 2004; Salvaterra, Haardt, Ferrara 2005). 
We show that future space-borne missions will be able to constrain different proposed 
models for the accretion history of MBHs at early times.  We focus here on a specific model for MBH formation,  which traces MBH seeds to the first generation of metal--free stars. We then compare how different accretion histories reflect onto the detectability of X--ray sources. 

The mass growth of the most massive BHs at high redshift must proceed very efficiently to explain
 the luminosity function
of luminous quasars at $z\approx 6$ in the Sloan Digital Sky Survey (SDSS, e.g. Fan 
et al. 2001). Volonteri \& Rees (2005) explore the conditions which allow a sufficient 
growth of MBHs by $z=6$ under the assumption that accretion is triggered by 
major mergers.  At such high redshift an investigation of evolution of the 
MBH population must take into account the dynamical evolution of MBHs.  
Black hole mergers, in fact, can give a net negative contribution to the early black 
hole growth, due to gravitational effects which can kick BHs out of 
their host halos (gravitational recoil, due to to the non-zero net linear 
momentum carried away by gravitational waves). 
Adopting recent estimates for the recoil velocity (Baker et al. 2006), 
Volonteri \& Rees (2005) find that 
if accretion is always limited by the Eddington rate via a thin disc,  the 
maximum radiative efficiency allowed to reproduce the LF at $z=6$ is 
$\epsilon_{\rm max}=0.12$ (corresponding 
to an upper limit to the MBH spin parameter of 0.8). If, instead, 
high-redshift MBHs can accrete at super-critical rate during an early phase 
(Volonteri \& Rees 2005, Begelman, Volonteri \& Rees 2006), then reproducing the 
observed MBH mass values, is not an issue. 
The constraints from the LF at $z=6$ are still very weak, so either a model with a 
low $\epsilon_{\rm max}$ or a model 
with supercritical accretion cannot be ruled out based on these results only. 

In a previous paper we have investigated how current observations can 
constrain the late accretion history of MBHs, at $z<4$ (Volonteri, Salvaterra 
\& Haardt 2006). In this paper we study by means of a detailed model of MBH 
assembly, the detectability of very high redshift AGN both at X--ray and 
near-IR wavelengths by future missions. We discuss the contribution of these objects to 
the unresolved  X--ray background (in particular in the 0.5--2 keV band). 
Finally, we
discuss the role of an early population of AGN in the reionization of the
Universe and their contribution to the UV background at high redshift.
The paper is organized as follows. In Sect.~2 we describe briefly the merger
tree model for the formation and evolution of MBH in the early Universe.
Sect.~3 presents the basic equations we need in the number counts and 
background calculation and, in Sect.~4, we discuss the adopted spectrum of
AGN. In Sect.~5 we present our results and discuss the role of future
space mission in the observations of high redshift AGN. In Sect.~6 we
compare our results to those of a model in which the early
evolution of MBH seeds is characterized by a phase of super--critical
accretion. Finally, in Sect.~7, we briefly summarize our results.

Throughout the paper we use the AB magnitude system\footnote{AB magnitudes are
defined as AB$=-2.5\log (F_{\nu_{0}}) -48.6$, where $F_{\nu_{0}}$ is the 
spectral energy density within a given passband in units of erg s$^{-1}$ 
cm$^{-2}$ Hz$^{-1}$} and the standard $\Lambda$CDM cosmology 
(Spergel et al. 2003).

\section{High redshift Massive Black Hole evolution}

The main features of a plausible scenario for the hierarchical
assembly, growth, and dynamics of MBHs in a $\Lambda$CDM cosmology
have been discussed in Volonteri, Haardt \& Madau (2003).  Dark matter halos 
and their associated galaxies undergo many mergers as mass is assembled from
high redshift to the present. The halo merger history is tracked
backwards in time with a Monte Carlo algorithm based on the extended
Press-Schechter formalism. ``Seed" holes are assumed to form with intermediate 
masses in the rare high $\nu-\sigma$ peaks  collapsing at $z=20-25$
(Madau \& Rees 2001) as end-product of the very first generation of stars.
As our fiducial model we take $\nu=4$ at $z=24$, which ensures that galaxies today 
hosted in halos with mass  larger than $10^{11}\;\Msun$ are seeded with a MBH. 
The assumed threshold allows efficient formation of MBHs in the range of halo
 masses effectively probed by dynamical studies of MBH hosts in the local 
universe. 

As a reference, we adopt here a conservative model assuming Eddington-limited accretion 
which is able to reproduce the bright end of the optical LF, as traced by observations in the SDSS (e.g. Fan et al 2001).  We then discuss in Section 6 a simple model which considers super-Eddington 
accretion rates for high-redshift MBHs and one in which more massive seeds form late, as in Koushiappas et al. (2004), 
and evolve through Eddington--limited accretion.

The fraction of active BHs (i.e., AGN) at early times is large in our models, reaching almost unity for the most massive black holes ($M_{\rm BH}>5\times 10^8 \Msun$) at $z\simeq 5$. The fraction of AGN with respect to the whole BH population decreases by at least one order of magnitude at $M_{BH} \sim 10^5-10^6\,M_\odot$ and further at lower masses.  Under our model assumptions, detection of active sources can therefore provide robust constraints on the whole BH population.

\section{Basic Equations}

The number of sources observed per unit solid angle at redshift $z_0$  in the flux range
$F_{\nu_0}$ and $F_{\nu_0}+dF_{\nu_0}$ at frequency
$\nu_0$  is 

%%%%%%%%%%%%%%%
\begin{equation}
\frac{dN}{d\Omega dF_{\nu_0}}(F_{\nu_0},z_0)=\int^{\infty}_{z_0}
\left( \frac{dV_c}{dz d\Omega} \right) n_c(z, F_{\nu_0})  \, dz,
\end{equation}
%%%%%%%%%%%%%%%

\noindent
where $dV_c/dz d\Omega$ is the comoving volume element per unit redshift per
unit solid angle, and $n_c(z,F_{\nu_0})$ is the comoving density of sources at
redshift $z$ with observed flux in the range $[F_{\nu_0}, F_{\nu_0} + dF_{\nu_0}]$.
The specific flux of a source observed at $z_0$ is given by
%%%%%%%%%%%%%%%%%
\begin{equation}\label{eq:flux}
F_{\nu_0}=\frac{1}{4\pi \, d_L(z)^2} \tilde L_\nu (M_{\rm BH}) \, \mbox{e}^{-\tau_{\rm eff}(\nu_0,z_0,z)},
\end{equation}
%%%%%%%%%%%%%%%%%
\noindent
where $\tilde L_\nu(M_{\rm BH})$ is the specific luminosity of the source (in units of erg 
s$^{-1}$ Hz$^{-1}$) {\it averaged over the source lifetime}, 
which is assumed to be only a function of the mass of the central BH. 
In the above eq.~(2), $\nu=\nu_0 (1+z)/(1+z_0)$, $d_L(z)$ is the 
luminosity distance, and 
$\tau_{\rm eff}$ is the effective optical depth of the intergalactic medium (IGM) 
at $\nu_0$ between redshifts $z_0$ and $z$. 
Shortward of the observed Lyman$-\alpha$, the emergent spectrum is strongly modified by IGM absorption (see Sect.~2.2 of 
Salvaterra \& Ferrara 2003 for a full description of the IGM modeling).  
For sources at $z>6$, as those we are considering in this paper, IGM absorption in the observed 0.5--10  keV band (corresponding, for sources at $z>6$ as those considered here, 
to rest--frame energy $>3.5$ keV) can be safely neglected.

The radiation background $J_{\nu_0}(z_0)$ observed at redshift 
$z_0$ at frequency $\nu_0$, is 
%%%%%%%%%%%% 
\begin{equation}\label{eq:J}
J_{\nu_0}(z_0)= \frac{(1+z_0)^3}{4\pi}\int^{\infty}_{z_0}
\epsilon_\nu(z) \mbox{e}^{-\tau_{\rm eff}(\nu_0,z_0,z)} \frac{dl}{dz}dz,
\end{equation}
%%%%%%%%%%%% 
where $\epsilon_\nu(z)$ is the comoving specific emissivity, and $dl/dz$ is the proper line element.
The source term $\epsilon_\nu$ is given by
%%%%%%%%%%%%
\begin{eqnarray}\label{eq:eps}
\epsilon_\nu(t)&=&\int dM_{\rm BH}\,\, \int_0^t L_{\nu}(t-t',M_{\rm BH})\frac{dn_c}{dt'dM_{\rm BH}}dt' \nonumber \\
&\simeq& \int dM_{\rm BH}\,\, \tau \tilde L_{\nu}(M_{\rm BH}) \frac{dn_c}{dt dM_{\rm BH}}. 
\end{eqnarray}
%%%%%%%%%%%%
The second approximated equality holds once we consider the  
source light curve averaged over the typical source lifetime $\tau$,  
and assuming the source formation rate per unit mass as constant over such timescale.

\section{AGN spectrum}

The physical characterization of the source is encoded in its spectral energy
distribution (SED), $\tilde L_{\nu}$. We model the UV part of the SED of unabsorbed AGN ($\log (N_{\rm H}/{\rm cm}^{-2})<22$, referred as type I) as a multicolor disk black 
body (Shakura \& Sunyaev 1973). Assuming  Eddington limited accretion, the 
maximum disk temperature is 
$kT_{\rm max} \approx 1\mbox{ keV}(M_{BH}/\Msun)^{-1/4}$. The characteristic 
multicolor disk spectrum is broadly peaked at 
$E_{\rm peak}\approx 3 kT_{\rm max}$,  
and follows a power law with $\tilde L_{\nu}\propto \nu^{1/3}$ for energies 
$h\nu \lta E_{\rm peak}$, and exponentially rolls off for $E \gta E_{\rm peak}$. 
In the X--ray, the spectrum can be described by a power--law with photon index
$\Gamma=1.9$ ($f_{\nu}\propto \nu^{-\Gamma}$), 
and an exponential cutoff at $E_c=500$ keV (Marconi et al.
2004). The averaged X--ray SED of absorbed AGN (type II) is described by the same type I spectrum for 
$E>30$ keV, and, in the range $0.5--30$ keV, 
by a power--law (continuously matched) with photon index $\Gamma=0.2$, obtained convolving the type~I spectrum with 
a lognormal distribution of absorption column density centered at $\log (N_H/cm^{-2})=24$  (Sazonov et al. 2004).
UV emission from type~II AGN is assumed to be negligible.   
We further assume a type~I/type~II ratio of 1/4, independently of redshift and
luminosity, since detailed modelling of intrinsic absorption at $z>6$
are not currently available. Note that the observed X--rays correspond, 
for sources at $z\simgt 6$, to rest--frame energies between $3.5$ and $70$ keV, where  
the emission properties of type~I and type~II AGN are thought to be 
similar (the rare Compton thick sources are not considered here).

The X--ray emission of type~I AGN is normalized to the optical, adopting an optical--to--X rays energy index  
$\alpha_{\rm OX}=0.126 \log(L_{2500})+0.01z-2.311$ (see Eq. 5 of Steffen et
al. 2006), where $L_{2500}$ is the monochromatic luminosity at 
$\lambda=2500$ \AA (rest frame). By definition, 
the X--ray luminosity at the rest frame energy of 2 keV, $L_2$, is
$L_2=L_{2500} (\nu_{2}/\nu_{2500})^{-\alpha_{\rm OX}}$. 
The scaling of $\alpha_{\rm OX}$ with redshift and luminosity has been obtained by Steffen et al. (2006) 
combining data from
the SDSS, COMBO-17 and {\it Chandra} surveys. The final sample consists of 333 AGN extending out to
$z\sim 6$ spanning five decades in UV luminosity and four decades in X--ray luminosity. 
Since only a mild dependence on redshift is found, we extrapolate this result also to $z>6$. 
We further assume Eddington limited accretion (see Section 2).

\section{Results}

\subsection{X-ray number counts}

%%%%%%%%%%%%%%%%%%%%%%%%%%%%%%%%%%%%
\begin{figure}\label{fig:xcount}
\center{{
\epsfig{figure=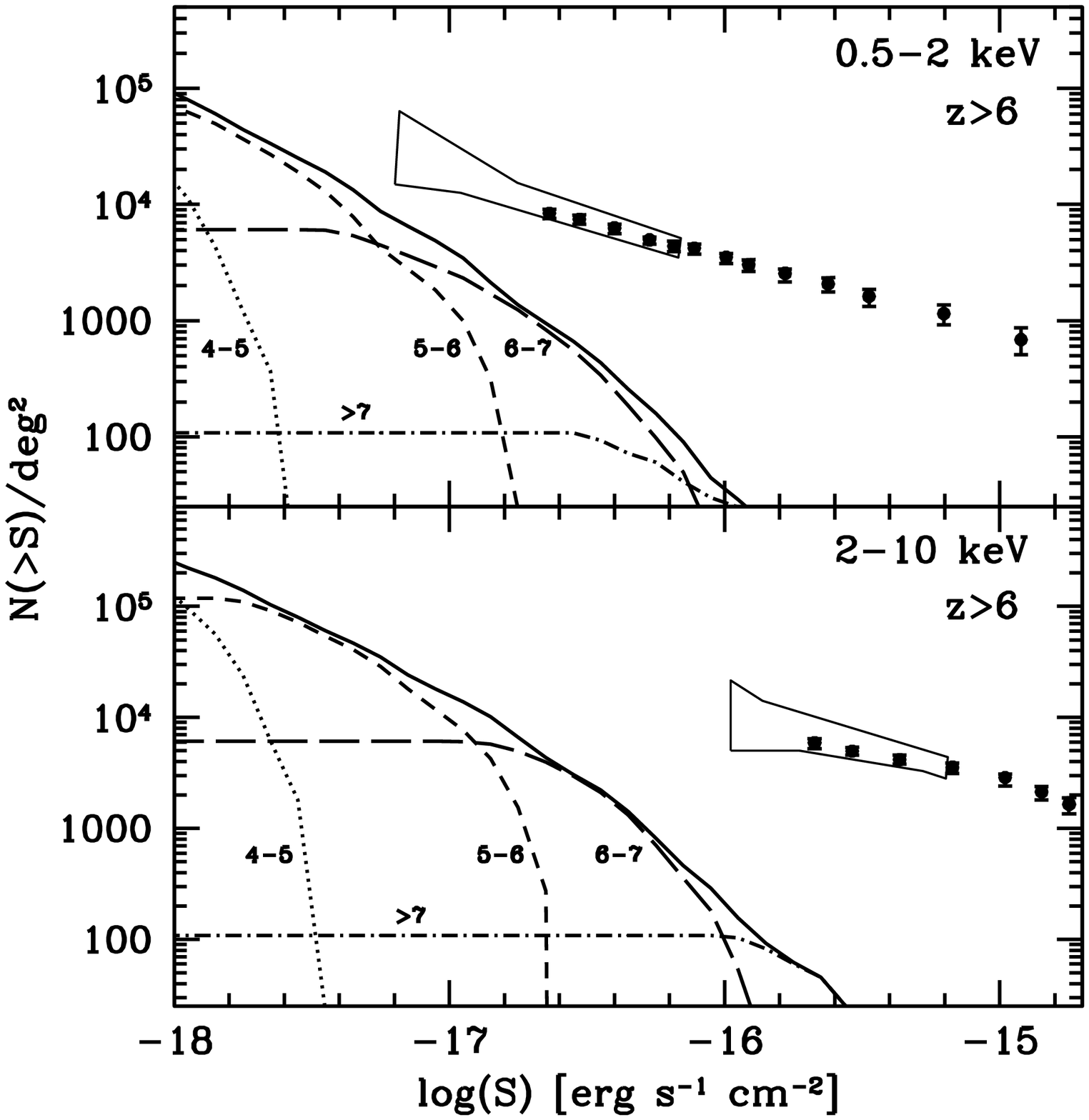,height=8cm} }}
\caption{
Predicted logN/logS in the soft ({\it upper panel}) and hard ({\it lower panel})
X--ray bands for sources at $z\simgt 6$.
Different lines refer to different BH mass ranges: all masses ({\it solid line}), 
$M_{\rm BH}=10^4-10^5\;\Msun$ ({\it dotted line}), 
$M_{\rm BH}=10^5-10^6\;\Msun$ ({\it dashed line}), 
$M_{\rm BH}=10^6-10^7\;\Msun$ ({\it long dashed line}), and 
$M_{\rm BH}>10^7\;\Msun$ ({\it dot--dashed line}). 
Data points are the measured source counts obtained by {\it Chandra} 
(Moretti et al. 2003) and the bow--tie indicates results from the fluctuation 
analysis (Miyaji \& Griffiths 2002).}
\end{figure}
%%%%%%%%%%%%%%%%%%%%%%%%%%%%%%%%%%%%

We compute the soft (0.5--2 keV) and hard (2--10 keV) X--ray number counts from MBHs shining at $z\simgt 6$ 
predicted by our model of MBH assembly and evolution. 
Results are shown in Fig.~1, and are compared to available 
observational data (Moretti et al. 2003). The bow--tie indicates results from the 
fluctuation analysis of the {\it Chandra} deep field (Miyaji \& Griffiths 2002). 
Different lines refer to different ranges in BH masses.
At the flux limit of current surveys, $\log(S)>-16.6 (-15.8) {\rm erg\,s^{-1}\,cm^{-2}}$  in the soft (hard) band (Alexander et al. 2003), 
the contribution to the logN/logS from sources at $z\simgt 6$ is $\sim 8\%$ in the soft X--rays, 
$\simlt 1\%$ in the hard band. 
In the 0.5-8 keV band, we expect $\sim 2$ source at $z\simgt 6$ in the {\it Chandra} Deep Field North with fluxes 
exceeding $3\times 10^{-16} \; \Xflux$, 
to be compared to an upper limit of  7 sources with extreme X--ray/optical flux ratios (EXO, Koekemoer et al. 2004),  that are candidate high redshift AGN.

The optical identification of these objects is problematic, owing to absorption
(both internal and by the IGM) and to their very low optical flux. 
Barger et al. (2003a, 2003b) searched for optical counterparts (at $5\sigma$ confidence) 
with $z$--band magnitude $z_{850}<25.2$ of AGN of the {\it Chandra} Deep Field North exposure. 
Apart from a source at $z=5.19$ with $z_{850}=23.9$, no other
$z>5$ candidate was identified. 
This is consistent with our results, where the majority of the
very high redshift sources detected in the deepest {\it Chandra} observations
should have a z--band magnitude fainter than 27. 
Barger et al. (2003b) used the lack of 
optical identifications to derive limits on the number of objects at
$z>5$. They found that in a field of view (f.o.v.) corresponding to 6 arcmin radius
circle, only $\sim 6$ sources with fluxes  exceeding $2\times 10^{-16}\;\Xflux$ in the 
0.4--6 keV band should lie at $z \gta 5$. Our model is consistent with 
this limit, predicting that only $\sim 1$ sources at $z\simgt 6$ should be 
found in the same field of view. 

At fainter fluxes, a significant fraction of the sources identified in
the fluctuation analysis of the deepest {\it Chandra} data might be AGN at 
$z\simeq 6$. 
Our model predicts that high redshift AGN contribute to the logN/logS  
at a level $\sim 13-60$\% in 
the soft X--ray band, $\lta 6$\% in the hard X--ray band at the flux limit of
the fluctuation analysis ($\simeq 8,000$ sources deg$^{-2}$ at the flux limit $\log S=-17.2$ for the soft band, 
and $\simeq 300$ sources deg$^{-2}$ at the flux limit $log S=-16$ for the hard band). 
Direct observations of such sources is
among the main goals of the next generation of X--ray telescopes (e.g. {\it
XEUS}\footnote{www.rssd.esa.int/index.php?project=XEUS} and 
{\it Constellation--X}\footnote{constellation.gsfc.nasa.gov}).
The {\it XEUS} mission is expected to have sufficient sensitivity to measure the 
X--ray spectra of sources as faint as $\sim 10^{-17}\;\Xflux$ in the 
0.5--2 keV energy range, while the photometric limiting sensitivity is expected to  
be $\sim 10^{-18}\;\Xflux$. 
In the hard--X band, the limiting sensitivity, both spectroscopic and photometric,  
will be larger by almost an order of magnitude. 
At the spectroscopic flux limit of {\it XEUS}, 
we predict almost $5\times 10^3$ ($300$) AGN in the soft (hard) X--ray band, 
within a 1 deg$^{2}$ f.o.v.. At such flux limits, $\sim 3\times 10^3$ ($\sim 85$) 
sources deg$^{-2}$ are type I objects, indicating that, because of obscuration, 
deep surveys in the soft (hard) X--ray band will miss nearly 90\% (20\%) of type~II AGN.   
At the photometric flux limits, we expect $\sim 10^5$ ($\sim 10^4$) sources deg$^{-2}$ in the soft (hard) 
X--ray band. In this case the type~II missing fraction is 70\% (7\%).

{\it XEUS} will be directly probing the lower end of the mass function of accreting MBHs at $z>6$, 
$M_{\rm BH}\sim 10^{5-6}\;\Msun$ (i.e. for luminosity $L_X>10^{42}$ erg s$^{-1}$ in the rest-frame 2--10 keV energy band). 
The main contribution to the number counts is still expected from sources at $z \lta 10$, but almost 
$10^3$ deg$^{-2}$ sources, i.e., 1\% of the sources, are expected to be detected at $z\gta 10$. 
We note here that, assuming an half--energy width of the PSF of 2", {\it XEUS} will be confusion limited 
at a sensitivity of $\simeq 4\times 10^{-18}\;\Xflux$, saturating the logN/logS at a level of 
$\simeq 2\times 10^5$ deg$^{-2}$ (Arnaud et al. 2000). 
For bright sources, redshift determination should be possible from the detection of X--ray emission lines (e.g., 6.4 keV K$\alpha$). For fainter sources, however, redshift determination in the 
X--rays might be very challenging, and a combination of deep X--ray observations and
ultra-deep optical/near-IR spectroscopy is probably required (see section 5.3).

In conclusion, we find that the next generation of X--ray missions will be able
to investigate the early stages of MBH build--up and to provide fundamental
information on the faint-end of the luminosity function of AGN even at very 
high redshift.

\subsection{X--ray background}

%%%%%%%%%%%%%%%%%%%%%%%%%%
\begin{figure}\label{fig:Xback}
\center{{
\epsfig{figure=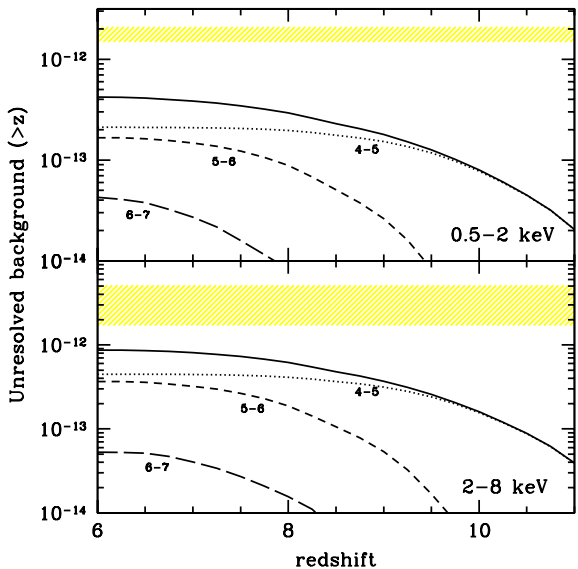,height=8cm} }}
\caption{
Cumulative contribution to the unresolved XRB (0.5--2 keV band, top panel and 2--8 keV band, bottom panel) 
as a function of redshift in $\Xback$. Line--style as in Fig. 1. 
The current observational limits (Hickox \& Markevitch 2006) 
are shown as a shaded area.
}
\end{figure}
%%%%%%%%%%%%%%%%%%%%%%%%%%

According to Moretti et al. (2003), the intensity of the total X--ray background (XRB) 
is  $7.53\pm 0.35 \times 10^{-12}$ and $2.02\pm 0.11 \times 10^{-11}\;\Xback$ in
the 0.5--2 keV, and 2--10 keV energy bands, respectively. 
A large fraction, $\simeq 94$\%, of the XRB in the 0.5--2 keV band has been 
attributed to sources with fluxes exceeding $2.4\times 10^{-17}\;\Xflux$, 
while $\sim 89$\% of XRB in the 2--10 keV band is resolved into sources 
whose flux is $\geq 2.1\times 10^{-16}\;\Xflux$ (Moretti et al. 2003). 
More recently, Hickox \& Markevitch (2006) estimate the unaccounted fraction
of the XRB due to extragalactic unresolved sources as 
$1.77\pm 0.31\times 10^{-12}\;\Xback$ in the soft X--ray energy band 
(0.5--2 keV) and $3.4\pm 1.7\times 10^{-12}\;\Xback$ in the hard X--ray 
energy band (2--8 keV).

We compute the contribution to the residual unresolved XRB in the soft and hard 
(0.5--2 keV and 2--8 keV, respectively) energy band  
from the population of AGN predicted by our model to exist at $z\gta 6$. 
Note that almost all these sources are below  the source detection
limit used by Moretti et al. (2003). 
Results are shown in Fig.~2, where the cumulative contribution to the XRB 
of sources at redshift $>z$ is considered. 
Different line--styles refer to different MBH mass ranges, as in Fig.~1.
The integrate contribution to the soft XRB from AGN at $z\simgt 6$ is found to be 
$\sim 0.4\times 10^{-12}\;\Xback$, corresponding to $\sim 5$\% of the total, 
and to $\sim 23$\% of the unaccounted fraction. 
Sources with masses $<5\times 10^5\;\Msun$ give the largest 
contribution to the XRB, whereas  
MBHs with $M>5\times 10^6\;\Msun$ produce a negligible background. 
Moreover, the major contribution to the unresolved X--ray background is from
MBHs shining at $z>9$. 
In the 2--8 keV band, the XRB from unresolved, $z\simgt 6$ 
AGN is $\sim 0.87\times 10^{-12}\;\Xback$, corresponding to $\sim 5$\% of 
the total observed hard XRB, and to $\simeq 25$\% of the unresolved fraction. 
MBHs with masses $<10^6\;\Msun$ give the main 
contribution also in the hard band, indicating that the unresolved fraction of the XRB
can be used to constrain the faint--end of the X-ray luminosity function of 
AGN at very high redshift.

We find that AGN at $z>6$ contribute significantly to the 
unaccounted XRB, although their contribution is still well below the available 
constraints. Other faint unresolved X-ray sources at $z<6$ 
may contribute to the X-ray background, including galaxies, starbursts (e.g., Bauer et al. 2004), and 
a population of faint AGN (e.g., Volonteri, Salvaterra \& Haardt 2006; see Table 1).

It is interesting to note that {\it at least} $\sim 3-4$\% of the observed XRB (0.5-10 keV) 
will remain unaccounted  
even after {\it XEUS} observations will become available, as due to sources at $z>6$ 
below the flux detection limit.
Only even more sensitive X--ray observatories such as the proposed 
{\it Generation--X}\footnote{generation-x.gsfc.nasa.gov} will assess 
these extremely faint sources.

\subsection{{\it JWST} number counts}

%%%%%%%%%%%%%%%%%%%%%%%%%%%%%%%%%%%%
\begin{figure}\label{fig:JWST}
\center{{
\epsfig{figure=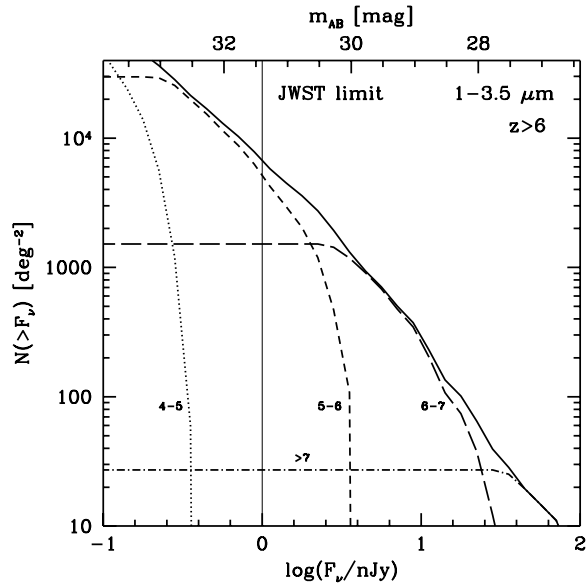,height=8cm} }}
\caption{Predicted number counts of AGN at $z\simgt 6$ in the $1-3.5\;\mu$m
band. Solid vertical line shows the expected {\it JWST} source detection limit.
Lines are the same of Fig. 1.}
\end{figure}

\begin{figure}\label{fig:uvback}
\center{{
\epsfig{figure=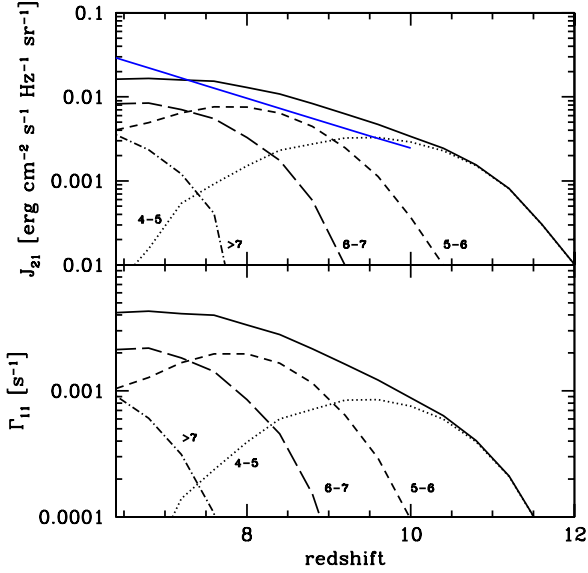,height=8cm} }}
\caption{Top panel: UV background at the Lyman Limit (912 \AA) as a function 
of redshift in units of $10^{-21}$ erg cm$^{-2}$ s$^{-1}$ Hz$^{-1}$ sr$^{-1}$.
 As reference value we show with the blue solid line the UV background as 
computed with the code CUBA (Haardt \& Madau 2006, in preparation) 
extrapolating the observed AGN luminosity function. 
Bottom panel: the corresponding HI ionization rate in units of 
$10^{-11}$ s$^{-1}$. Lines are the same of Fig. 1.}
\end{figure}
%%%%%%%%%%%%%%%%%%%%%%%%%%%%%%%%%%%%

Deep--field observations in the near-IR might be able, in principle,
to detect AGN at very high redshifts. 
The $i$--band ACS data in the Hubble Ultra Deep Field (HUDF) 
reach sensitivity as the 28th mag (Beckwith et al. 2006). 
Our model predicts that, at this limiting magnitude, the sky density
of AGN with $z\simgt 6$ is $\sim 0.03$ arcmin$^{-2}$. 
For comparison, 
the observed sky density of $z>6$ galaxies in the HUDF is $\sim 4.7$ arcmin$^{-2}$; Bouwens et 
al. 2005), corresponding to an AGN fraction of $\simeq 1\%$, so that the probability to detect these sources in a field as small 
as the HUDF ($\simeq 12$ arcmin$^2$) is extremely low. 
Given the actual size of the HUDF (Bouwens et al. 2005), 
an area three times larger has to be observed at the same magnitude limit in 
order to detect one AGN at $z>6$.
In order to have a statistically significant number of very
high redshift AGN, lower detection limits are needed. 
The James Webb Space Telescope ({\it JWST}\footnote{ngst.gsfc.nasa.go}) is expected to 
detect sources down to a flux limit of $\sim 1$ nJy (i.e., $m_{AB} \sim 31.4$), 
in a f.o.v. of $4\times 4$ arcmin$^2$. 
We have computed the number counts predicted by our model for near-IR
deep-field observations. 
Results are shown in Fig.~3. At the expected sensitivity of
{\it JWST} almost $7\times 10^3$ deg$^{-2}$ sources at $z\simgt 6$ 
should be detected, down to a mass limit for the accreting MBHs of 
$\sim 3\times 10^5\;\Msun$, almost the same mass scale accessed by the deepest 
future X--ray surveys. 
In spite of this good result, we have to note that {\it JWST} 
observations will be able to reveal only the un--absorbed population of 
AGN, whereas X-ray survey will probe the entire AGN population. 
Moreover, the detection of high redshift AGN in the deepest {\it JWST} data 
could be hampered by the difficulty to select high-$z$ AGN from the more 
numerous population of high redshift galaxies. 

In conclusion, X--ray 
surveys are a better tool to investigate the early stage of MBH 
assembly, allowing to detect all (type I and type II) AGN to the highest 
redshifts and to the lowest mass range. {\it JWST} deep surveys of AGN 
might be useful to investigate the optical part of the SED, and the 
evolution of the type~I/type~II ratio at very high redshifts. 
However, optical/near-IR follow up (with {\it JWST} and the forthcoming 30 meter class telescopes) 
of X--ray selected faint sources can be used to redshift determination. 
Unobscured AGN should be recognizable both from the presence of broad emission lines
and distinctive optical-near-IR colours while obscured AGN should in general be identifiable
from the detection of high-excitation optical/near-IR emission lines.
For $z>6$ AGN the strongest line will be Lyman $\alpha$ at 1216 \AA. Further in the infrared CIV (1549 \AA), HeII (1640 \AA), CIII (1909 \AA) and MgII (2800 \AA), will be all detectable by JWST,  provided the gas has substantial metallicity.  Of these lines the CIV line is probably the strongest.
A combination of {\it JWST}, 30 meter class telescopes and deep X-ray observations
should provide the most effective identification strategy of AGN.

We have to note here that MBHs shining at high redshifts, similarly to miniquasars (Salvaterra, Haardt
\& Ferrara 2005), are not expected to contribute 
significantly to the cosmic background in the near-IR, 
and do not represent a viable solution of the near-IR background excess problem 
(Salvaterra \& Ferrara 2003, 2006).

\subsection{UV background and reionization}

We have also computed (see eqs. 3 and 4) the background intensity at the 
Lyman Limit (912 \AA) as a function of $z_{0}$ using the opacity of 
Fardal, Giroux \& Shull (1998; model A1).
Results are shown in Fig.~4. 
As a reference, we show the UV background (blue solid line) 
as computed with an updated 
version of the code CUBA (Haardt \& Madau 1996), extrapolating the 
observed AGN luminosity function and IGM opacity. 
We find a good agreement between the two 
different approaches. In the bottom panel of Fig.~4 
we show the corresponding HI ionization rates.
The UV background is dominated by the most massive, i.e. 
$>10^6\;\Msun$, MBHs up to $z\simeq 7$. At higher redshifts, these objects are rarer, 
and the background is mostly due to MBHs in the mass range $10^5-10^6\;\Msun$.
At $z>10$ the background intensity falls rapidly.

%%%%%%%%%%%%%%%%%%%%%%%%%%%%%%%%%%%%
\begin{figure}\label{fig:reion}
\center{{
\epsfig{figure=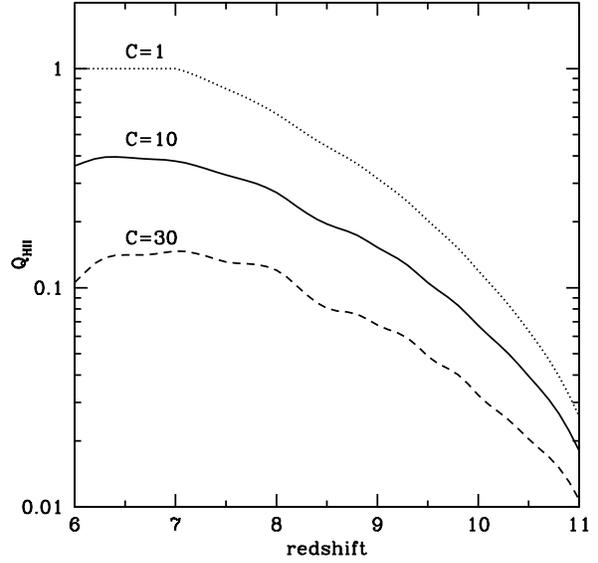,height=8cm} }}
\caption{Redshift evolution of the filling factor of HII regions, 
$Q_{\rm HII}$, for different values of the clumping factor: $C=1$ (dotted
line), $C=10$ (solid line), and $C=30$ (dashed line). Complete reionization 
is reached when $Q_{\rm HII}=1$.}
\end{figure}
%%%%%%%%%%%%%%%%%%%%%%%%%%%%%%%%%%%%

Finally, we may ask whether high redshift AGN can contribute significantly
to the reionization of the Universe. In order to answer this question, we 
compute the redshift evolution of the filling factor of HII regions as
(Barkana \& Loeb 2001),

\begin{equation}
Q_{\rm HII}(z) = \int_{z}^{\infty} dz^\prime
\left\vert\frac{dt}{dz}\right\vert \frac{1}{n_H^0}
\frac{dn_{\gamma}}{dt} \mbox{e}^{F(z^\prime,z)},
\label{eq:reio}
\end{equation}

\noindent
where $n_H^0 = X_H n_B^0$ and $n_B^0$ are the present-day number densities of
hydrogen and baryons ($X_H=0.76$ is the hydrogen mass fraction), and
$dn_{\gamma}/dt$ is the production rate of ionizing photons. The function 
$F(z',z)$ takes into account the
effect of recombinations. Assuming a time-independent
volume-averaged clumping factor $C$, common to all HII regions, 
we can write, 

\begin{equation}
F(z^\prime,z) = -\frac{2}{3} \frac{\alpha_B n_H^0}{\sqrt{\Omega_M} H_0} C 
[f(z^\prime)-f(z)],
\end{equation}

\noindent
and 

\begin{equation}
f(z)=\sqrt{(1+z)^3 + \frac{1-\Omega_M}{\Omega_M}},
\end{equation}

\noindent
where $\alpha_B = 2.6 \times 10^{-13} \mbox{cm}^3 \mbox{s}^{-1}$ is
the hydrogen recombination rate. 

The evolution of $Q_{\rm HII}$ as function of redshift is shown in Fig.~5 for
$C=1$ (dotted line), $C=10$ (solid line) and $C=30$ (dashed line). For $C=1$,
AGN alone are able to reionize completely the Universe at $z\sim 7$. 
For higher values of $C$, although the model fails to reach $Q_{\rm HII}=1$, 
we find that the AGN contribution is not negligible and might 
help in substaining the ionization of the IGM at $z\sim 6$.

\section{Comparison with rapid--growth model}

%%%%%%%%%%%%%%%%%%%%%%%%%%%%%%%%%%%%
\begin{figure}\label{fig:LFX}
\center{{
\epsfig{figure=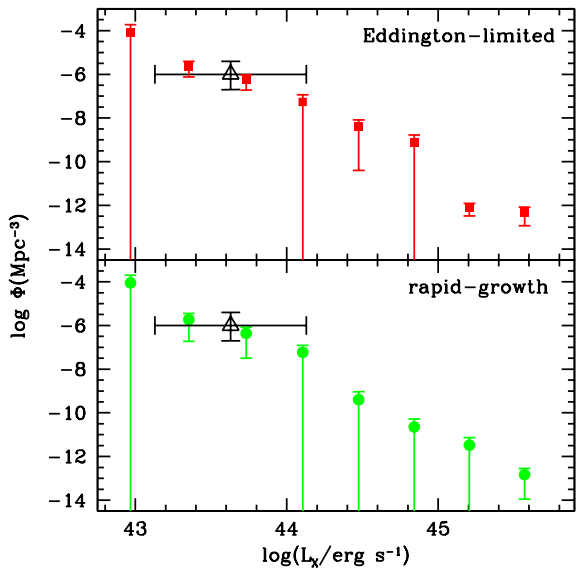,height=8cm} }}
\caption{Predicted luminosity function of quasar in the rest--frame hard X--ray 
band [2--10 keV] at $z=6$. The open triangle shows the estimated number density
of quasars in the {\it Chandra} Deep Field North (Barger et al. 2003b). 
Top panel shows the result for the Eddington--limited
model, whereas the bottom panel shows the result  for the rapid--growth model.}
\end{figure}
%%%%%%%%%%%%%%%%%%%%%%%%%%%%%%%%%%%%%

In this section, we discuss possible differences between the Eddington--limited
model and a model allowing MBHs to accrete at super--critical rate during 
the early phases of their evolution. In Fig.~6 we plot the LF in the 
rest--frame hard X--ray band [2-10 keV] at $z=6$ for the two models (top panel:
Eddington--limited model; bottom panel: rapid--growth model). Since
both models reproduce the observed optical LF at $z=6$ (Volonteri \& Rees 2005,
Begelman et al. 2006),  the X--ray LFs are also quite similar. For the
Eddington--limited model the differential X--ray LF can be described by a 
simple power--law with power index $\beta=-4.55$, whereas for the 
super--critical accretion model 
$\beta=-4.77$. The open triangle in Fig.~6 shows The estimated number density 
of $z\sim 5.7$ AGN as obtained by Barger et al. (2003b) in the 
{\it Chandra} Deep Field North. Both models are in good agreement with the
observed number density.

Although the two models share similar results at $z=6$, the
LF at higher redshift shows significant differences. For
example, the number density of bright quasars at $z=10$ (i.e. with luminosity
in the rest-frame hard X--ray band in the range $5\times 10^{42}<L_X <
3\times 10^{43}$ erg s$^{-1}$, corresponding to MBHs with $10^6<M_{BH}<10^7\;\Msun$)
is $\sim 10^{-5}$ Mpc$^{-3}$ for the super--critical accretion model, whereas in the
Eddington--limited model this density decreases by almost an order of magnitude.
This result is confirmed by Fig.~7, where the solid line shows the redshift 
distribution of sources with fluxes above $10^{-17}\;\Xflux$ (the planned 
spectroscopic flux limit of future X--ray missions) in the observed soft X--ray
band for the two models. A model that allows an early phase of
super--critical accretion will have a redshift distribution of sources
pushed towards higher redshifts with respect to the Eddington--limited model. Dotted 
lines in Fig.~7 show the redshift distribution of sources with X--ray fluxes
above $10^{-17}\;\Xflux$ that may also be observable with future near-IR facilities 
such as {\it JWST} (i.e., at the source detection limit of $F_\nu \sim 1$ nJy), showing that future X--ray 
observations along with optical (near-IR) identifications by the next generation of
space and 30--meter class telescopes might be able to discriminate the two different 
accretion histories. 
Also shown in Fig.~7 is the redshift distribution of sources with fluxes
above $10^{-16}\;\Xflux$, easily achieved by deep {\it Chandra} observations (e.g., Alexander et al. 2003). 
Furthermore, we note here that at the planned soft X--ray photometric 
limit of {\it XEUS}, only
$\simeq 1$\% of observable sources is expected to lie at $z\simgt 10$, if BHs  
growth is Eddington--limited, whereas this fraction 
increases up to $\simeq 18$\% in case of super--critical accretion.

It is unlikely that sufficient $z>6$ AGN will be identified by the Chandra deep field observations to discriminate between these two models. However, we can provide indirect constraints by considering the contribution that each make to the unresolved X-ray background. We find
that the contribution to the unresolved XRB increases by a factor $\sim 3$: 
the predicted unresolved background from sources at $z\simgt 6$ is  
$\sim 1.2\times 10^{-12}\;\Xback$ in
the 0.5--2 keV band, and $\sim 2.3\times 10^{-12}\;\Xback$ in
the 2--8 keV band. Results are summarized in table~1. 
As a net result, once that the contribute to the unresolved XRB of sources at $z<4$ is considered, 
rapid--growth model saturates the observed unaccounted background. 
Note that, according to Worsley et al. (2005), most of the unaccounted 2--8 keV XRB 
actually lies at energies $\simgt 6$ keV, strengthening our conclusions.

We also find that the X--ray background is saturated in a model 
in which massive seeds form late, as in Koushiappas et al. (2004). In this model seed MBH  form from the low angular momentum tail of material in halos with efficient gas cooling.   In first approximation, the Koushiappas et al. (2004) model indicates that seed MBH form in halos with mass above the threshold $M_H\simeq10^7 M_\odot (1+z/18)^{-3/2}$, with a mass $m_{\rm seed}\simeq5\times10^4 M_\odot(M_H/10^7 M_\odot)(1+z/18)^{3/2}$.  The largest seed mass allows for a less efficient accretion, in particular, we have assumed that the accretion rate is  Eddington--limited, and the maximum radiative efficiency is 20\% (Gammie et al 2004).   Also in the late formation scenario, the unresolved XRB is 
saturated, being $\sim 1.5\times 10^{-12}\;\Xback$ in
the 0.5--2 keV band, and $\sim 3.2\times 10^{-12}\;\Xback$ in
the 2--8 keV band.
XRB saturation could be a problem for such models, as a substantial contribution to the XRB from low--redshift (i.e. 
$z<4$), unresolved faint sources (Volonteri, Salvaterra \& Haardt 2006) and/or 
galaxies (Bauer et al. 2004) can not be excluded.

\begin{table}\tiny
\begin{minipage}{0.5\textwidth}
\begin{center}
\begin{tabular}{lccl}
\hline
\hline
Sources & \multicolumn{2}{c}{XRB} & Reference \\
 & 0.5--2 keV & 2--8 keV & \\
\hline
Faint AGN ($z<4$) & 0.7 & 3.5 & Volonteri et al. 2006\footnote{model IIIb} \\
galaxies & 
0.4\footnote{based on extrapolation down to $\log{F_\nu}=-18$ in c.g.s units}
& 0.2\footnote{based on extrapolation down to $\log{F_\nu}=-17$ in c.g.s units} 
& Bauer et al. 2004 \\
\hline
\multicolumn{4}{c}{AGNs at $z>6$} \\
\hline
Eddington limited & 0.4 & 0.9 & this paper \\
Rapid growth & 1.2 & 2.3 & this paper  \\
Massive seeds & 1.5 & 3.2  & this paper \\
\hline
\multicolumn{4}{c}{Observed unresolved XRB} \\
\hline
 & 1.77$\pm$0.31 & 3.4$\pm$1.7 & Hickox \& Markevitch 2006 \\
\hline
\end{tabular}
\end{center}
\caption{Contribution to the unresolved XRB from different sources in
units of $10^{-12}\;\Xback$.}
\end{minipage}
\label{tab:xrb}
\end{table}

%%%%%%%%%%%%%%%%%%%%%%%%%%%%%%%%%%%%%
\begin{figure}\label{fig:zdist}
\center{{
\epsfig{figure=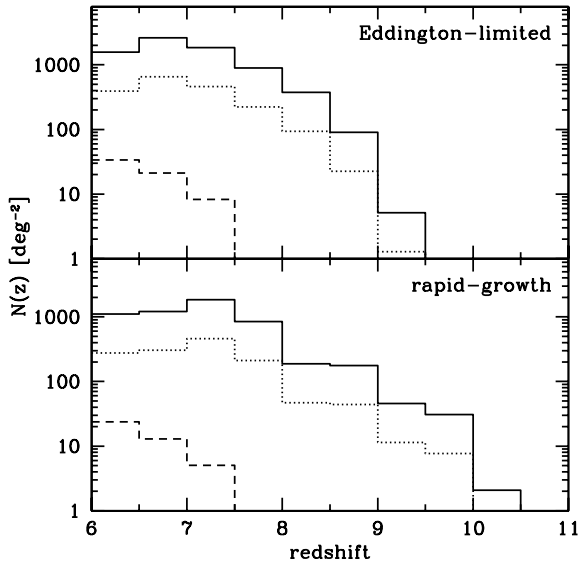,height=8cm} }}
\caption{Redshift distribution of sources with fluxes in the observed soft
X--ray band [0.5--2 keV] above $10^{-17}\;\Xflux$ (i.e., the 
planned spectroscopic flux limit of future X--ray missions, solid line), and above
$10^{-16}\;\Xflux$ (easily achieved by {\it Chadra} deep field observations, dashed line). 
Dotted line the ones with near-IR fluxes above the planned {\it JWST} 
sensitivity. Top panel shows the result for the Eddington--limited
model, whereas bottom panel the one for the rapid--growth model (see Section 6).}
\end{figure}
%%%%%%%%%%%%%%%%%%%%%%%%%%%%%%%%%%%%

\section{Summary and Conclusions}

In this paper we have assessed, using Monte Carlo simulations of DM halo merger history coupled with 
semi--analytical recipes for the assembly of MBHs within galaxy spheroids (Volonteri et al. 2003), 
the possibility of constraining the AGN population
at $z\simgt 6$ with currently  available and planned space borne missions. 
In particular, we have considered ultra--deep X--ray and near-IR surveys. 
We claim that, among the unresolved sources in {\it Chandra} deep fields, few of them are 
AGN at $z\simgt 6$, whose BH masses are $\simgt 10^7\; \Msun$, while in
the near-IR, an are three times larger than the HUDF area is required in order 
to detect just one of these high--z sources. 

Future X--ray missions,  such as {\it XEUS}, and near-IR  
facilities such as {\it JWST}, will have the technical capabilities to detect accreting MBHs 
at $z\simgt 6$ down to a mass limit as low as $10^5-10^6\;\Msun$. 
Since {\it JWST} observations will reveal only the un--absorbed AGN population, X--ray deep 
surveys are definitely the best tool suited for investigation of the early stages of MBH assembly. 
In particular, the next generation of X--ray telescopes might detect
as many as $\sim 10^5$ sources/deg$^{2}$ at their photometric sensitivity 
limits in the 0.5--2 keV band. 
Further constraints on the population of high redshift AGN will be 
provided by future 21cm experiments like LOFAR (Rhook \& Haehnelt 2006).

We have shown that our predicted population of high redshift AGN  
would account for a significant fraction of the unresolved XRB (0.5--8 keV). Almost 5\%
of the measured XRB (or $\sim 25$\% of the unresolved one) should come
from sources at $z\simgt 6$. Moreover, in our model of $z>6$ AGN, the major contribution  
actually comes from sources at $z>9$, with fluxes 
$<10^{-18}\;\Xflux$, i.e. that can not be detected even by the next generation
of X--ray telescopes; we find that at least 3-4\% of the measured XRB should
remain unresolved even after {\it XEUS} observations become available. 
These constraints become much more severe for a model in which super--critical 
accretion is allowed in the early stages of the MBH growth (Volonteri \& Rees
2005). The model reproduces well the observational constraints at 
$z=6$ (optical and X--ray LFs), and predicts a more extended tail 
of sources observable at $z\simgt 9$. However, we find that in the rapid--growth model, 
the predicted XRB is three times higher than in the Eddington--limited accretion model, saturating the 
unresolved fraction of the XRB in both the 0.5--2 keV and 2--8 keV energy bands. 
Since faint sources at $z<4$ are expected to contribute substantially to the 
unaccounted XRB (Bauer et al. 2004; Volonteri et al. 2006), this result suggests that the 
occurrence and effectivity of super-critical accretion should be investigated 
in much more detail. In particular, super--Eddington accretion could be much less efficient in few--$\sigma$
peaks halos, due to the gas evacuation from the ionizing radiation emitted by the MBH seed Pop III star progenitor (Johnson \& Bromm 2006). Super--Eddington accretion consequently is biased towards the highest density peaks, which experience the largest number of mergers with halos containing pristine gas to replenish the gas reservoir. 
Models in which seeds are much more massive than Pop III star remnants, as in Koushiappas et al. (2004), 
saturate the unresolved fraction of the XRB in both the 0.5--2 keV and 2--8 keV energy bands as well. 
We note here that different models for MBH seed formation, although predicting rather large BH masses, can have 
a lower formation efficiency, which can ease the constraints given by the XRB (see Eisenstein \& Loeb 1995, 
and Begelman, Volonteri \& Rees 2006). 
Our constraints on theoretical models are conservative, as we adopted the largest value of the 
unresolved XRB available in literature.

Finally, we have computed the evolution of the  
UV background produced by the modeled population of high redshift AGN. 
Later than $z\simeq 7$, the ionizing intensity is dominated by relatively massive BHs, $M\gta 10^6\;\Msun$, 
while lighter BHs contribute mostly at earlier epochs. 
The UV background from AGN rapidly declines at
$z\gta 10$. We compute the contribution of these sources to the reionization
of the Universe, showing that AGN alone can provide enough ionizing photons only in the (improbable) case 
of an almost completely homogeneous inter--galactic medium. Nevertheless, for a more clumpy medium, 
the AGN contribution to the ionizing background is never negligible.

We note that high redshift AGN can not contribute significantly to the 
near-IR background.

\section*{acknowledgments}
We wish to thank A. Ferrara for fruitful discussions on reionization issues, and 
the referee, David Alexander, for thoughful comments that improved the quality of the paper. This research was supported in part by the National Science Foundation under Grant No. PHY99-07949. NSF-KITP-06-89 pre-print.

\end{document}